\documentclass{mem}
\usepackage{natbib}\usepackage{txfonts}\usepackage{balance}
\usepackage{flushend}
\usepackage{graphicx}
\usepackage[breaklinks,pdftex]{hyperref}

\begin{document}

\title{
Gamma-ray astronomy from the ground – future perspectives
}


\author{
J.A. \,Hinton\inst{1} 
} 
\institute{Max-Planck-Institut f\"ur Kernphysik, Saupfercheckweg 1, D-69117 Heidelberg, Germany\\
\email{jim.hinton@mpi-hd.mpg.de}\\
}
\authorrunning{Hinton}

\titlerunning{Gamma-rays from the ground – future perspectives}

\date{Received: 13-02-2025; Accepted: 28-08-2025}

\abstract{
I provide a personal perspective on the future of the field of ground-based gamma-ray astronomy, on the occasion of the 2024 {\it Gamma} conference in Milan. I discuss some of the scientific motivations for new instrumentation and the major new projects that are in development or already under construction, together with emerging concepts for instrumentation in the farther future. I stress the strong complementarity of the ground-level particle detector arrays, with their wide-field capabilities, and the more precise Cherenkov telescope arrays. The key science topics for the next decades require both approaches and both are developing rapidly towards major performance advances and full sky coverage. I will briefly outline the status and roles of the projects CTAO and SWGO which will dominate the next decade. Beyond these projects are several developments which might boost performance at both ends of the ground-based gamma-ray energy range, including the plenoscope approach at low energies and diverse approaches to ultra-high-energy gamma-ray astronomy; from lake-based instruments to arrays of very small Cherenkov telescopes. I will again briefly review these activities and how they may contribute long term. 
}
\maketitle{}

\section{Introduction}

By conventional definition the gamma-ray domain in astrophysics covers approximately 12 of the 23 explored orders of magnitude of the electromagnetic spectrum. From this huge range the highest five orders of magnitude from tens of GeV to beyond PeV, can be measured from the ground. Ground-based astronomy has huge advantages with respect to launching dedicated satellites, and -- when the required collection area exceeds tens of m$^{2}$ -- it is unavoidable. Ground-based approaches make use of the electromagnetic cascades (air-showers) initiated in the Earth's atmosphere by the primary photons. Two strongly complementary techniques have now been established: measurement of air shower particles at ground level and imaging of emitted air-Cherenkov light from showers with arrays of telescopes.
The Imaging Atmospheric Cherenkov Telescopes (IACTs) were the first successful approach, with the detection of TeV emission from the Crab Nebula in 1989~\citep{WhippleCrab}. Arrays of telescopes quickly followed and the leading instruments of the current generation: HESS, VERITAS and MAGIC, were established around 2004.

The development of the ground-level particle detectors came somewhat later, with the first major successes around 2004 with the MILAGRO instrument~\citep{milagro_gps}, followed by HAWC since 2015~\citep{hawc_nim} and LHAASO since 2020~\citep{lhaaso_km2a_crab}. The breakthrough for this type of detector came from improved background rejection, ultimately related to large ground coverage for capturing muons and electromagnetic sub-showers, with respect to earlier instruments.

The two approaches complement each other very strongly. The Cherenkov telescopes provide access to the lowest energies, with large collection area, and the greatest precision in both arrival direction and energy reconstruction for primary gamma-rays, in dedicated pointed observations during darkness. The ground-level particle detectors in contrast provide a continuous view of the whole overhead sky, with very large exposures possible at ultra-high energies, but with limited precision and low energy capabilities.

The wide range of scientific motivations for new instrumentation based on these two approaches is discussed in Section~\ref{sec:motivations} below, followed by a brief status report on the two major new facilities (CTAO and SWGO) in Section~\ref{sec:observatories}. Finally, emerging concepts for the future of the ground-based astronomy field are discussed in Section~\ref{sec:concepts}.


\begin{figure*}
    \centering
    \includegraphics[width=1.0\linewidth]{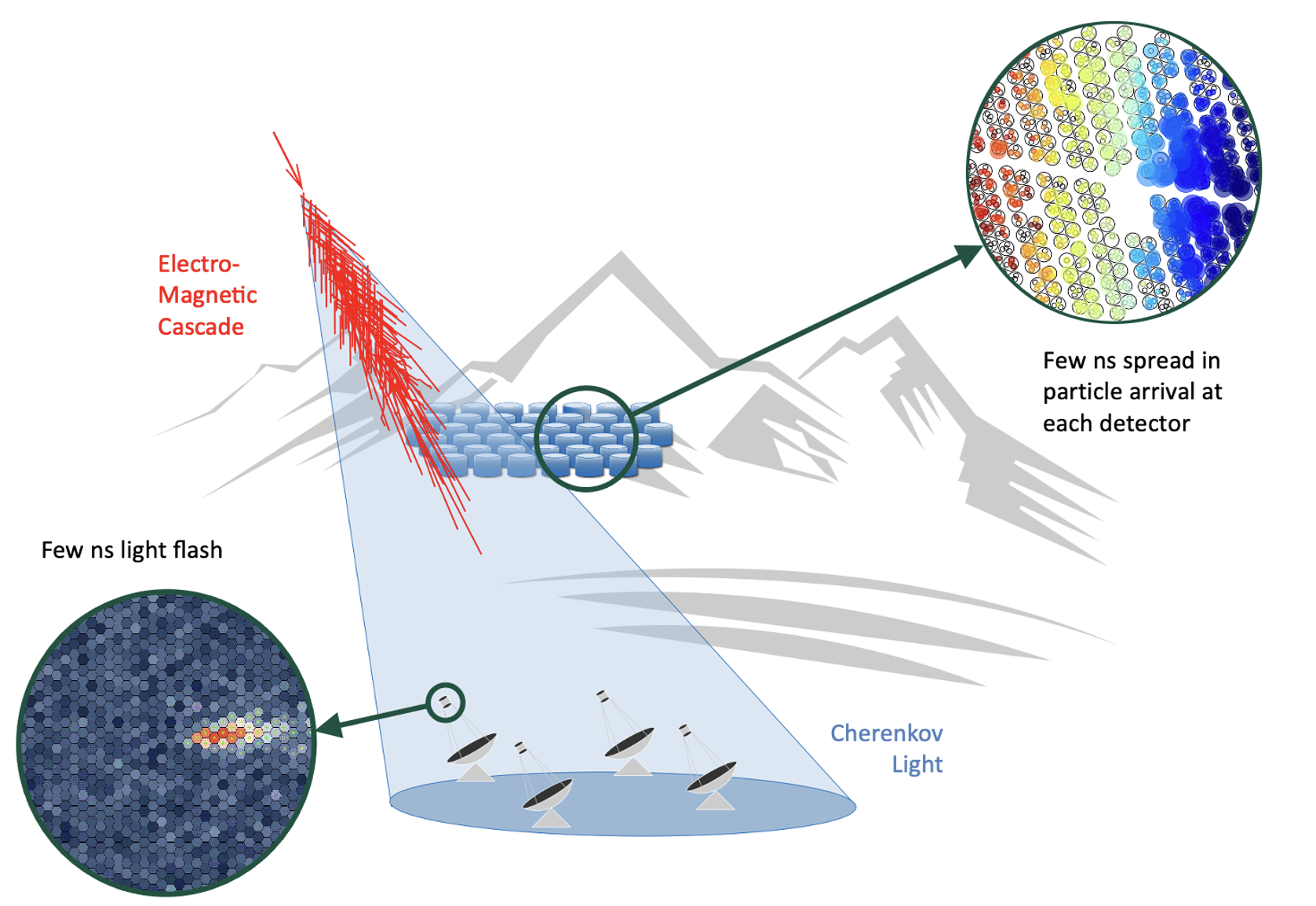}
    \caption{Illustration of the two approaches to ground-based gamma-ray astronomy: the imaging atmospheric Cherenkov technique and the ground-level particle technique. Adapted from~\cite{armelle_thesis}.    
    \label{fig:sketch}}
\end{figure*}

\begin{figure*}[t!]
\resizebox{\hsize}{!}{\includegraphics[clip=true]{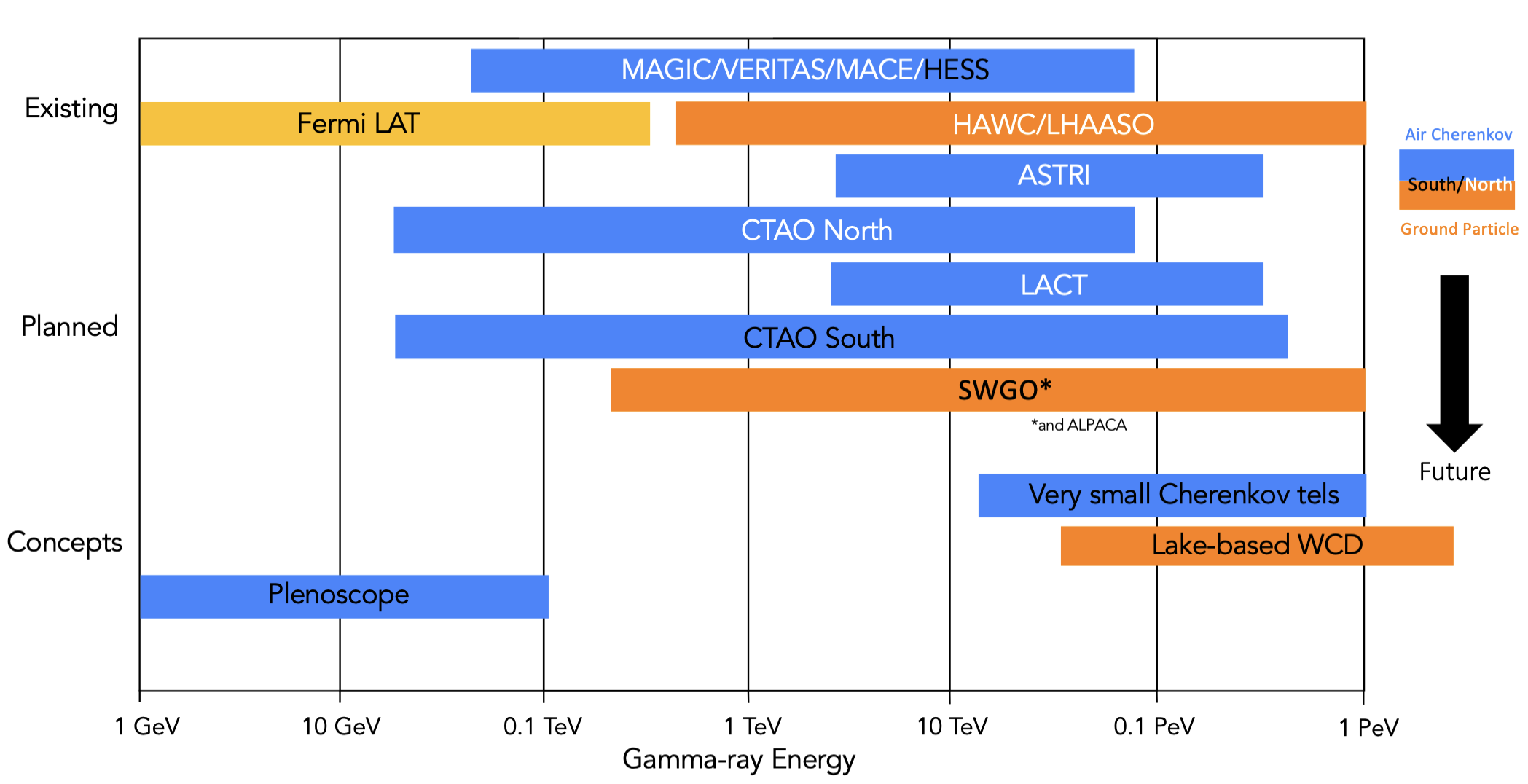}}
\caption{\footnotesize
Current and near future projects for ground-based gamma-ray astronomy, together with concepts for future development. The projects are organised by approximate maturity / timescale for realisation and indicative energy ranges. Instruments based on ground-level particle detection (orange bars) are distinguished from those utilising air-Cherenkov telescopes (blue bars). The location of the instruments is indicated by the text colour with Southern hemisphere instruments in black.
}
\label{fig:future}
\end{figure*}

\section{Scientific motivations for new instrumentation}
\label{sec:motivations}

In this section I give a small selection of the motivational science topics for new generation instrumentation, covering topics in high energy astrophysics (Section~\ref{sec:pevatrons}), multi-messenger astronomy (Section~\ref{sec:vary}) and beyond standard model (particle) physics (Section~\ref{sec:wimp}).

\subsection{Pevatrons: a tale of two hemispheres}
\label{sec:pevatrons}

Identification of the sites and mechanisms of particle acceleration up to PeV energies in our own galaxy has been a long-standing astrophysical challenge. Isolated supernova (SN) remnants, once the most widely favoured object class to provide the bulk of the cosmic rays up to the knee, have encountered a number of theoretical and observational challenges. Emerging alternatives include massive stellar clusters (or SN exploding inside their parent clusters) and microquasars \citep[see e.g.][]{westerlund1, thibault, ss433}. 

Thanks to the HAWC and (in particular) LHASSO observatories the Northern sky, also the home of the well-established IACT arrays MAGIC and VERITAS, is now well studied at energies on to PeV~\citep[see e.g.][]{lhaasopevsources}.
In stark contrast the southern hemisphere has a single major instrument: H.E.S.S., and no wide-field/ground-level particle instrument.

\begin{figure}
    \centering
    \includegraphics[width=1.0\linewidth]{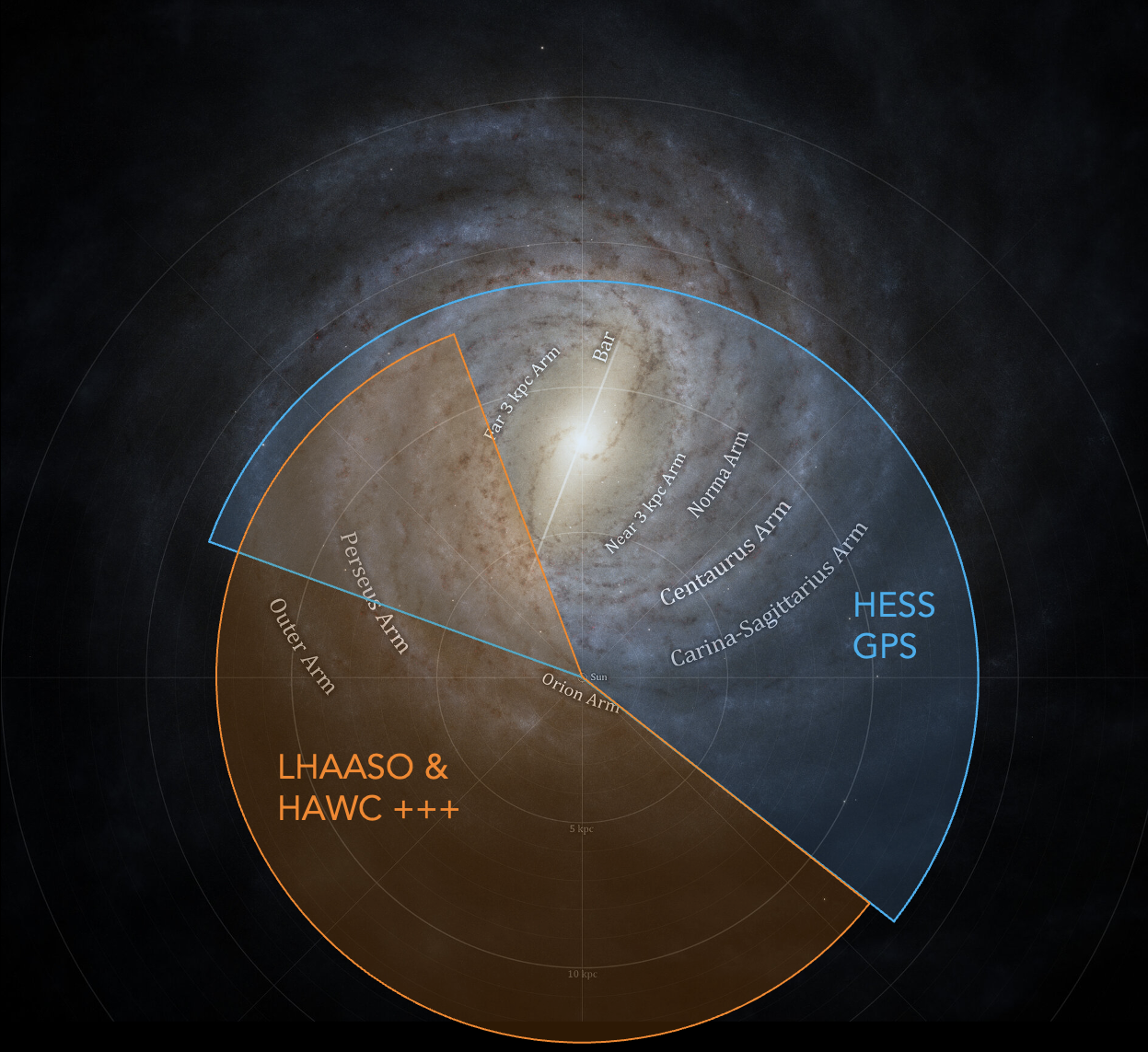}
    \caption{Approximate coverage of the Milky Way by ground-based gamma-ray surveys. Background image from ESA/Gaia/DPAC, Stefan Payne-Wardenaar, CC BY-SA 4.0 IGO.}
    \label{fig:surveys}
\end{figure}

This is not just half of the sky (see Section~\ref{sec:vary}),
but the bulk of our own galaxy that is not yet covered at UHE, as illustrated in Fig.~\ref{fig:surveys}. The southern sky contains most of the candidate objects for acceleration to PeV energies and of course the Galactic Centre itself, with evidence of close to PeV particles injected in to the Central Molecular Zone~\citep{hesspevgc}.

A combination of the very large exposure on individual objects possible with ground-level particle detectors, and the 
unchallenged angular resolution of the IACT arrays, is needed in the southern hemisphere to make progress.

\subsection{The rapidly varying gamma-ray sky}
\label{sec:vary}

The instantaneous wide-field capabilities of the ground-particle detectors are obviously of great interest for transient and time variable phenomena, but conversely the IACTs have superior short-timescale performance and greater cosmological reach. A combination of the two approaches has great promise in this area.

Unveiling the ultra-fast variability of TeV blazars was one of the highlights of current generation IACTs \citep[see e.g.][]{pks2155, magicflare}, indicating very high bulk Lorenz factors and extremely compact emission regions. Except in really exceptional states, and for the closest objects (due to EBL absorption) these minute timescales are out of reach for the ground-level particle detectors, but they can provide effective monitoring of bright objects on week and month timescales, as triggers for the IACTs \citep[see e.g.][]{hawcalert}.

Truly transient phenomena are a relatively recent addition to the TeV sky, with gamma-ray burst detections from the ground only since 2018. The detections with MAGIC and HESS contained many surprises, including hard spectra beyond TeV energies and emission lasting deep into the afterglow phase~\citep{hessafterglow}.

So far the only GRB detected with a ground-particle array is the remarkable 'BOAT' GRB 221009A~\citep{lhassoBOAT}. The huge statistical significance and very wide energy range of the LHAASO measurement were able to provide detailed physical constraints. This event was far from typical of course, but it does demonstrate the promise of combined observations of ground-level particle arrays and IACTs. In general the promise of multi-wavelength and multi-messenger observations of transient phenomena in the years to come is very strong \citep[see e.g.][]{JimEdna}.

\subsection{A thermal relic of the big bang}
\label{sec:wimp}

A weakly interacting massive particle (WIMP), left over as a {\it thermal relic} of the big bang, remains as one of the most compelling candidates for the Dark Matter (DM) that makes up 26\% of our universe. In such a scenario the (velocity-weighted) self-interaction cross-section of the WIMP is essentially fixed by the current dark matter density. This gives indirect searches (looking for signatures of DM interactions in astrophysical environments) a distinct advantage in comparison to direct searches, which probe the less-constrained WIMP-nucleon cross-section. Of course there are also disadvantages of the indirect approach, uncertainties on the DM profile in the inner halos of galaxies are typically high, and astrophysical foregrounds are a significant complication, but non-the-less there is huge potential for a detection of WIMP annihilation in the halo of our own galaxy.

As the brightest annihilation signature is expected from the Galactic Centre, the key to this scientific target is reaching the
critical level of sensitivity in the southern hemisphere.
The combination of high resolution and sensitivity from IACTs with the wide field and UHE sensitivity of ground level particle detectors is again extremely powerful for this science case \citep[see e.g.][]{aion_dm}.

\section{The major observatories}
\label{sec:observatories}

Two major new ground-based observatories are planned to be realised within the next $\sim$5 years: CTAO and SWGO. The Cherenkov Telescope Array Observatory is designed to have transformational impact on a broad range of science cases~\citep{cta_science} and has two sites to provide access to the whole sky.

CTAO North, located on the Canarian Island of La Palma, is well advanced. The construction of the first large-sized telescope, LST-1, was completed already in 2018 and the construction of the remaining three (23 m diameter) LSTs is well advanced (see Zanin et al., these proceedings). Nine additional medium-sized telescopes (MSTs, 12 m diameter) are planned to complete the Northern array, making CTAO North a powerful tool for observations of (in particular) the high redshift and transient universe.

At CTAO South, infrastructure deployment is underway and observations with the first telescopes are expected to begin in 2026. The anticipated \href{https://www.ctao.org/news/ctao-releases-layouts-for-alpha-configuration/}{'alpha' configuration} contains 14 MSTs and 37 small-sized telescopes (SSTs, 4~m diameter). High energy events are seen with very large telescope multiplicities at CTAO South, providing excellent angular resolution (in particular when coupled to  the latest shower reconstruction algorithms, \citep[see e.g.][]{freepact}), reaching below an arcminute at tens of TeV, totally unprecedented in the field of gamma-ray astronomy. A key science project for CTAO is a deep survey of the Galactic Plane~\citep{cta_gps}, with the southern array playing the predominant role. See Zanin et al. (these proceedings) for more information.

Only 330~km away from CTAO South is the chosen site of SWGO: Pampa La Bola in the Atacama Astronomical Park. The location allows for the coverage of all transient events occurring above CTAO South to be monitored with SWGO, with detections or constraints from SWGO in the time window where CTAO telescopes are being alerted and/or slewing. At 4.8~km above sea level SWGO will have unprecedented low-energy reach for a ground-level particle detector, and aims for LHAASO-like UHE capabilities, with an outer array on a 1~km$^{2}$ scale. With the galactic centre passing directly overhead, SWGO will probe the PeV emission of the central region, the nature of the Fermi bubbles, and together with CTAO South, has the potential to confirm or rule out a thermal relic WIMP over a very wide mass range (see Ren et al., these proceedings).

The SWGO collaboration is currently preparing a modest scale pathfinder array of Water Cherenkov Detectors (WCDs) to be deployed over the next year. Major funding applications are in preparation around the world, to support the next construction steps beyond the pathfinder (see Barres de Almeida et al., these proceedings).

\begin{figure*}
    \centering
    \includegraphics[width=0.9\linewidth]{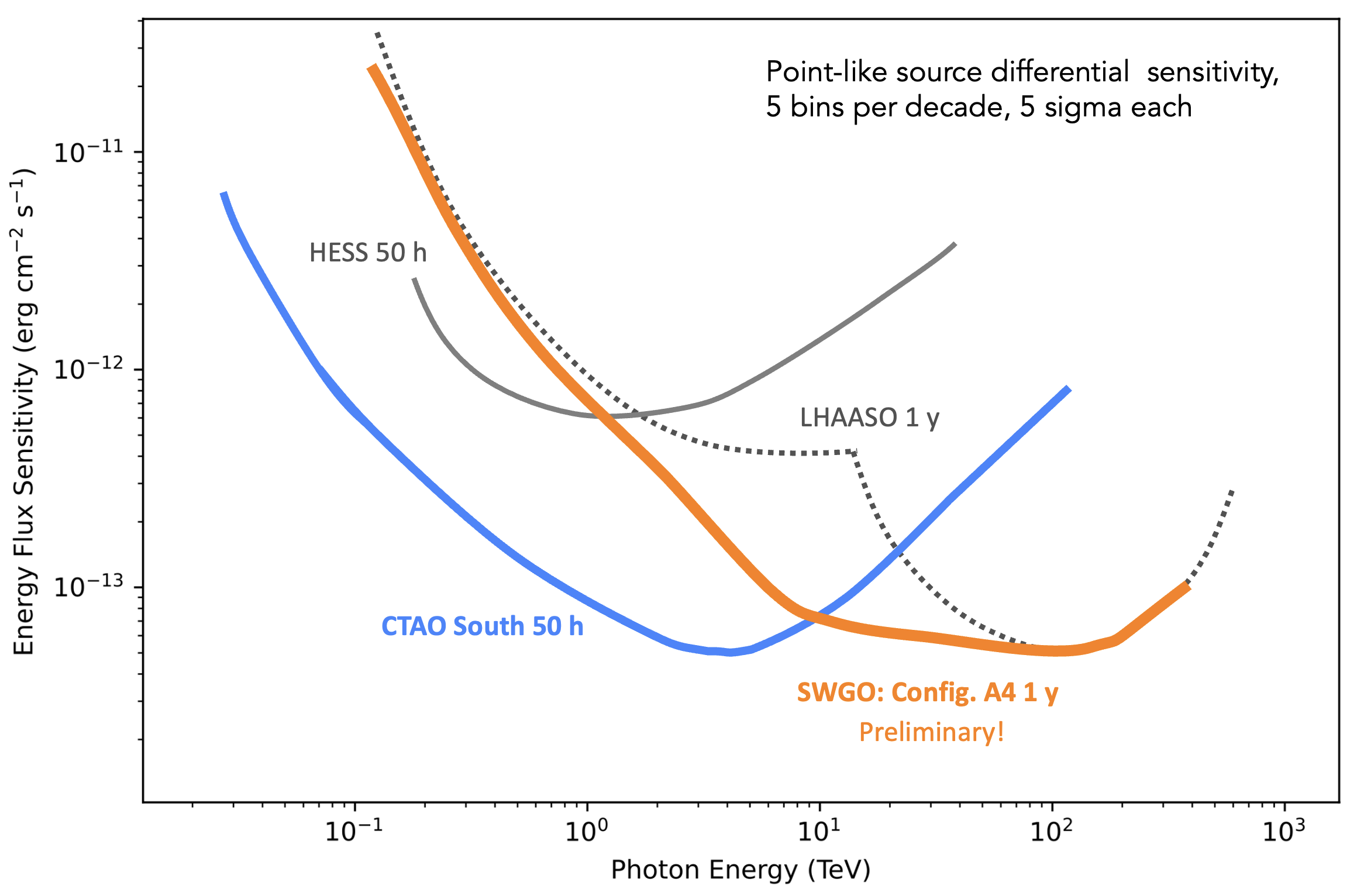}
    \caption{Sensitivity comparison of CTAO and SWGO to existing instruments HESS and LHAASO, in terms of differential sensitivity to point-like gamma-ray sources.}
    \label{fig:perf}
\end{figure*}

Fig.~\ref{fig:perf} gives an indication of the expected performance for CTAO South and SWGO in comparison to LHAASO and HESS, for point-like sources. Science-case studies including SWGO and CTAO are in preparation for an upcoming SWGO collaboration publication.

\section{Developing concepts}
\label{sec:concepts}

\subsection{The Plenoscope}

One clear opportunity for the ground-based future is to push in to the energy range below that of CTA, motivated in part by the exploration of very short timescale phenomena, in to the distant universe.
The ground-based detection of showers in the 1--10 GeV domain is possible in principle but extremely challenging for a number of reasons, including
triggering on low light levels, reconstructing primary parameters given large fluctuations in shower development, and background rejection (for similar reasons).

One approach is an array of (modest sized) telescopes exchanging low-level information to allow triggering on showers well below the threshold of an individual telescope \cite[see e.g.][]{triggersct}. The alternative would be
a single very large telescope that somehow overcomes the restrictions of depth of field effects. In fact this is not possible with a telescope, but requires a {\it plenoscope}, which registers the impact position of photons on the dish as well as their angles, using a light-field camera.
This option has been developed by M\"uller and co-workers~\citep{plenoscope}, and would allow
offline re-focusing and correction of aberrations, greatly easing the mechanical tolerances in supporting the primary mirror.

Such an approach becomes more attractive as costs for compact photosensors and associated readout electronics fall \citep[see e.g.][]{chec-s}, but clearly a 70~m single dish will still represent a very serious investment.  
The science-performance of such an instrument (in particular its expected background rate) is under investigation, complicated by the extremely strong geomagnetic effects which come in to play at such low energies.

\subsection{Very small IACTs}

The existence of low cost Cherenkov cameras with physically small (3-6 mm) pixels opens up a new possibility also at ultra-high-energies: 0.5-1~m diameter telescopes in large numbers. The most developed concept is PANOSETI~\citep{panoseti}, which has already been tested at the VERITAS site.

\begin{figure*}
    \centering
    \includegraphics[width=0.514\linewidth]{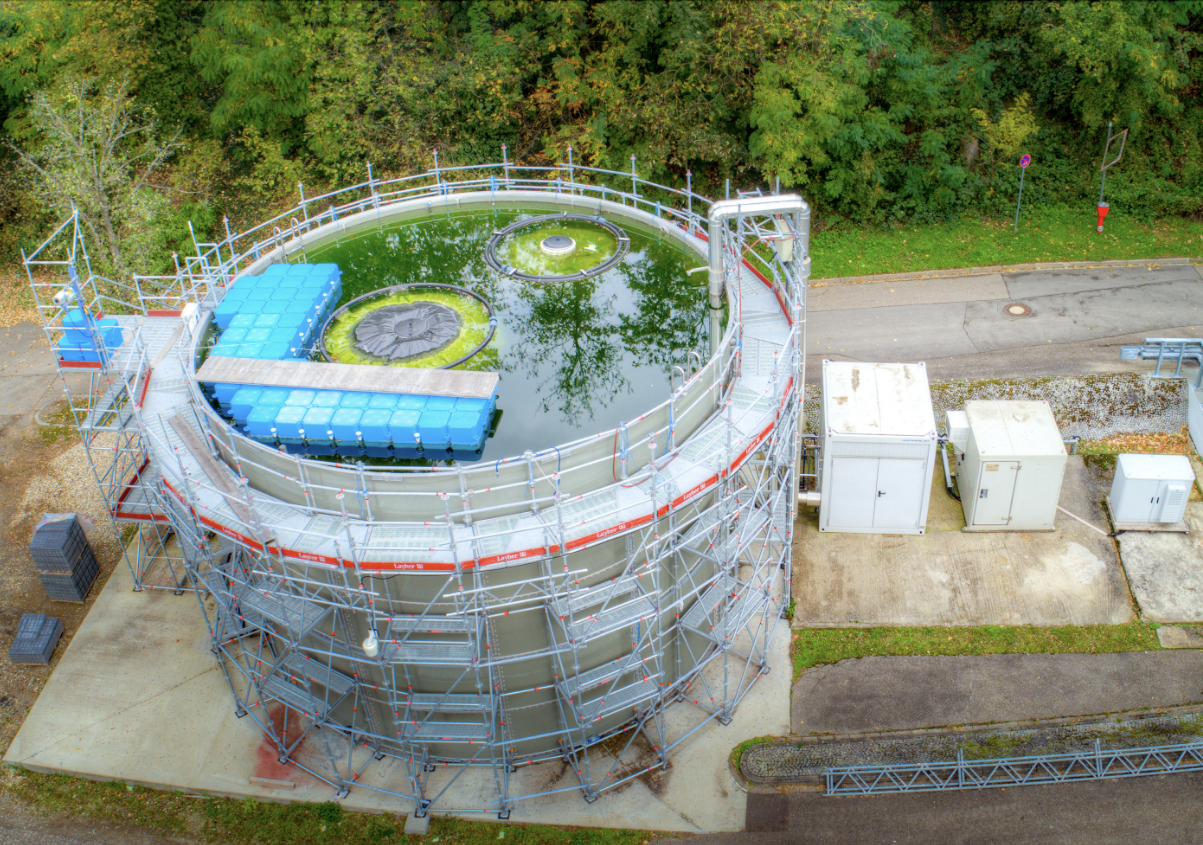}\hspace{0.5mm}\includegraphics[width=0.481\linewidth]{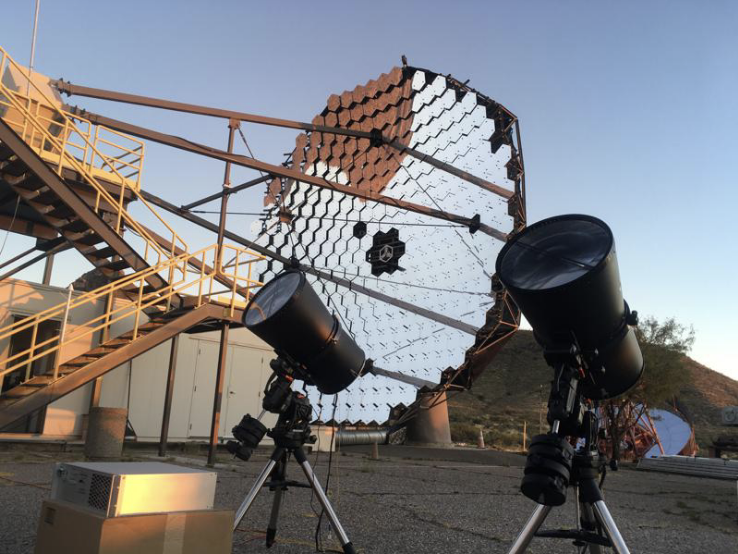}
    \caption{Left: lake simulation tank at MPIK, Heidelberg, with deployed prototype WCDs (reproduced from~\citet{lake_icrc}). Right: PANOSETI telescope in-front of VERITAS telescope T4 (reproduced from~\citet{panoseti}).}
    \label{fig:uhe}
\end{figure*}

The telescopes were development to perform a {\it Panoramic Search for Extraterrestrial Intelligence}, with sensitivity to nanosecond timescale optical flashes, and are hence well suited to Cherenkov observations. Commerical telescope mounts are used to point the 0.5~m diameter Fresnel lens which focus light on to 32$\times$32 pixel SiPM cameras (see Fig.~\ref{fig:uhe} (right)).

Such an approach can surely be used to obtain reasonable energy and angular resolution at multi-100 TeV energies, the main challenge is background rejection at the level required for deep observations of extended objects. Hence, it remains to be proven if this can be competitive with ground-level particle arrays as a cost-effective method for UHE astronomy. 

Hybrid systems of course can be very interesting - with background rejection via muon tagging in a surface array, and the Cherenkov telescopes supporting event reconstruction. However, it is important to bear in mind that the fraction of coincident events in such a hybrid system will always be low, given the limited FoV and duty cycle of the air-Cherenkov component.

\subsection{The lake-based UHE array}

The idea of a lake-based detector for neutrinos or high-energy single muons, and hence deep under water, is well established, in particular at Lake Baikal~\citep{baikal}. A surface-based lake array can however be interesting for gamma-ray astronomy and has been investigated in the context of SWGO~\citep{lake_icrc,lake_paper}. The attraction of such a detector is that water containers (tanks etc) can be avoided, and in particular in the context of muon detection in the style of LHAASO KM2A~\citep{lhaaso_km2a_crab}, digging of large holes and pouring of concrete etc. becomes unnecessary.
Prototyping work on such detectors is underway at MPIK in Germany and also in China, 
including directly at the site of LHAASO. An artificial lake was built at MPIK to test deployment concept and operation of bladders, see Fig.~\ref{fig:uhe} (left).

After selection of a land-based site for SWGO, the consideration of a lake-based option is focussed on a possible future UHE extension. Simulations and cost estimation are underway, and critically: the search for a suitable site.

As the high-altitude lakes of the world are typically of great environmental and cultural importance, very sensitive site-selection is a must. The deployed bladders have intrinsically low environmental impact (they are filled with pure water and are designed to keep this water pure long term), but possible disturbance to wildlife and local communities must be carefully considered.






\section{Conclusions}

The construction of CTAO proceeding on La Palma and the advanced plans for CTAO-South and SWGO make this a very exciting time for ground-based gamma-ray astronomy, with the prospect of a stream of major discoveries in astrophysics and astroparticle physics in the years to come. In parallel the development of several new concepts provides fertile ground for the development of the field for a long time to come.

\begin{acknowledgements}
I thank my colleagues in CTAO and SWGO for their suggestions for these proceedings, as well as the excellent collaboration.
\end{acknowledgements}

\bibliographystyle{aa}
\bibliography{bibliography}

\end{document}